\documentclass[10pt]{article}

\usepackage{graphicx}
\usepackage{float}
\usepackage{amsmath}
\usepackage{amsfonts}
\usepackage{jheppub}
\usepackage{multirow}
\usepackage{colortbl}
\definecolor{kugray5}{RGB}{224,224,224}
\usepackage{color}

\def\tb{\bar{t}}

\def\spaa#1.#2.#3{\langle\mskip-1mu{#1}|#2|{#3}\mskip-1mu\rangle}
\def\spbb#1.#2.#3{[\mskip-1mu{#1}|#2|{#3}\mskip-1mu]}
\def\spa#1.#2{\left\langle#1\,#2\right\rangle}
\def\spb#1.#2{\left[#1\,#2\right]}
\def\spab#1.#2.#3{\left\langle#1|#2|#3\right]}
\def\spba#1.#2.#3{\left[#1|#2|#3\right\rangle}

\def\P3b{\bar{P}_3}

\def\g0{\gamma_0}

\newcommand{\beq}{\begin{equation}}
\newcommand{\eeq}{\end{equation}}
\newcommand{\beqn}{\begin{eqnarray}}
\newcommand{\eeqn}{\end{eqnarray}}

\newcommand{\slsh}{\rlap{$\;\!\!\not$}}     

\title{$t \tb\, W^\pm$ production and decay at NLO}

\author{
    John M. Campbell and R. Keith Ellis
    \\
    Fermilab, Batavia, IL 60510, USA
    \\
    E-mails: 
    {\tt johnmc@fnal.gov}, 
    {\tt ellis@fnal.gov}.}

\preprint{
FERMILAB-PUB-12-109-T}

\abstract{
We present results for the production of a top pair in association with a $W$-boson at
next-to-leading order.  
We have implemented this process into the parton-level integrator MCFM including the
decays of both the top quarks and the $W$-bosons with full spin correlations.
Although the cross section for this process is small, it is a Standard Model source of
same-sign lepton events that must be accounted for in many new physics searches.
For a particular analysis of same-sign lepton events in which $b$-quarks are also present,
we investigate the effect of the NLO corrections as a function of the signal region cuts.
}
\keywords{QCD, Hadron colliders, LHC}

\begin{document}

\maketitle

\section{Introduction} 
In this note we present results for the production of a top quark pair in association 
with a $W$-boson, calculated at next-to-leading order (NLO) in QCD.
The core of the calculation, performed at amplitude level, is essentially identical
to the calculation performed for the $b \bar{b} \, W$ process in ref.~\cite{Badger:2010mg},
where results were given retaining the mass of the $b$ quark.
Since the  calculation is performed at amplitude level we can treat the top quarks as on-shell but 
include the effects of decay, retaining all spin correlations~\cite{Campbell:2004ch,Campbell:2012uf}.
The decay is included using a generalization of the amplitude method described in 
ref.~\cite{Campbell:2012uf}.
The complete process has been implemented into MCFM v6.3.

\section{$t \tb\, W$ production without decay}

\begin{figure}
\begin{center}
\includegraphics[angle=270,width=10cm]{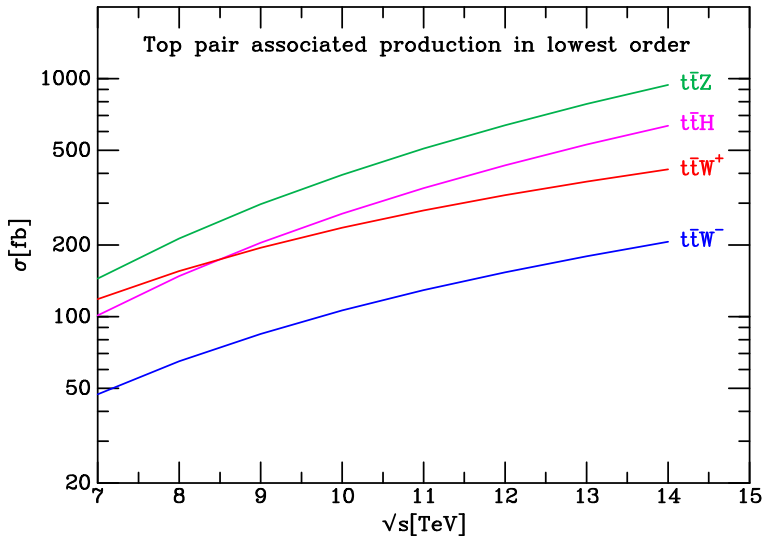}     
\caption{Lowest order cross sections for several processes in which a $t \tb$ pair
is produced in association with a boson at a $pp$ collider, as a function of the centre of mass energy.
Renormalization and factorization scales have been set equal to $m_t=172.5$~GeV and the $t\tb H$ process
is computed for a Higgs mass of $125$~GeV.}
\label{overview}
\end{center}
\end{figure}

To set the scene we display lowest order cross sections for several processes in which a $t \tb$ pair 
is produced in association with a massive boson in Fig.~\ref{overview}.
These are the total cross sections without branching ratios to observable modes and we consider
LHC operating energies from~$7$ to $14$~TeV. We note that, since we are presenting results for a proton-proton
collider, the rates for $t \tb\, W^+$ and $t\tb\, W^-$ are not equal.
As well as the $t \tb\, W$ process we have included the
LO rates for $t \tb\, Z$ and $t \tb\, H$, which are of a similar size.
At the current energy of the LHC, $\sqrt{s}=7,8$~TeV, the cross sections are quite small, so that measurements of 
these processes may have to wait for data produced at the maximum energy of the LHC machine.
Despite the fact that measurements have not yet been performed, these processes already play a role in searches for 
new physics. For this reason it is essential that the best possible theoretical predictions for these processes
are made available. Next-to-leading order corrections to the Higgs signal process at the LHC were presented in
Refs.~\cite{Beenakker:2002nc,Dawson:2002tg,Dawson:2003zu} and the $t\tb\, Z$ process has also been computed at this
level~\cite{Lazopoulos:2007bv,Lazopoulos:2008de,Kardos:2011na,Garzelli:2011is}.
Here we focus on results for the remaining $t\tb\, W$ process.

In passing we note that the production of other particles in association 
with a top quark pair has also been considered in the literature.
For example, NLO results for the production of $t \tb+$jet were presented in ref.~\cite{Melnikov:2011qx} 
and predictions for $t \tb\, \gamma$ were given in ref.~\cite{Melnikov:2011ta}. The $t \tb\,  b \bar{b}$ process
was also computed at NLO
in refs.~\cite{Bredenstein:2008zb,Bredenstein:2009aj,Bevilacqua:2009zn,Bredenstein:2010rs}.
We shall not comment further on these processes since they are not the focus of this paper.


Before proceeding to consider results for the $t \tb\, W$ cross section including the decay,
we present NLO results for the $t\tb\, W$ cross section for a stable top quark and top antiquark.
After adjusting the mass parameter, this calculation is essentially identical 
to our $ b \bar{b}\, W$ calculation presented in ref.~\cite{Badger:2010mg}.
We note that a value for the NLO rate for the process $t\tb\, W^+$ at $\sqrt{s}=7$~TeV 
has been given in Ref.~\cite{Hirschi:2011pa},
although to the best of our knowledge no detailed phenomenological studies have been performed.

Throughout this paper we use the following electroweak parameters,
\begin{equation}
M_W = 80.398~\mbox{GeV}\;, \;\; 
\Gamma_W = 2.1054~\mbox{GeV} \;,\;\;
G_F = 1.16639 \times 10^{-5} \, \mbox{GeV}^{-2}\;,
\end{equation}
and take the top and bottom quark pole masses to be,
\beq
m_t = 172.5~\mbox{GeV}\;, \;\; m_b = 4.7~\mbox{GeV} \;.
\eeq
For the parton distribution functions (pdfs) we use the sets of Martin, Stirling,
Thorne and Watt~\cite{Martin:2009iq}. For the calculation of the LO results presented here we employ
the LO pdf fit, with 1-loop running of the strong coupling and
$\alpha_s(M_Z)=0.13939$. Similarly, at NLO we use the NLO pdf fit, with
$\alpha_s(M_Z)=0.12018$ and 2-loop running.

\begin{figure}
\begin{center}
\includegraphics[angle=270,width=10cm]{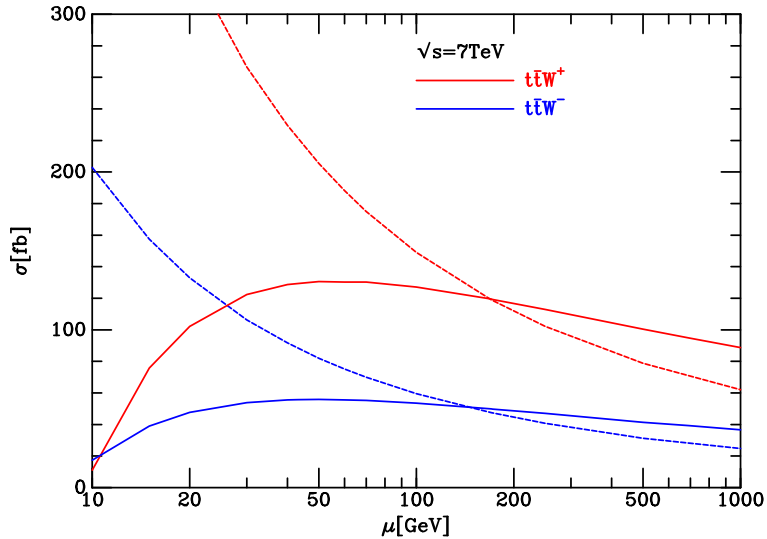}     
\caption{Dependence of the total $t \tb\, W^\pm$ cross sections on the renormalization and factorization scales
$\mu$ at $\sqrt{s}=7$~TeV. The scale dependence is reduced going from LO (dashed curves) to NLO (solid curves).}
\label{ttW7}
\end{center}
\end{figure}

\begin{figure}
\begin{center}
\includegraphics[angle=270,width=10cm]{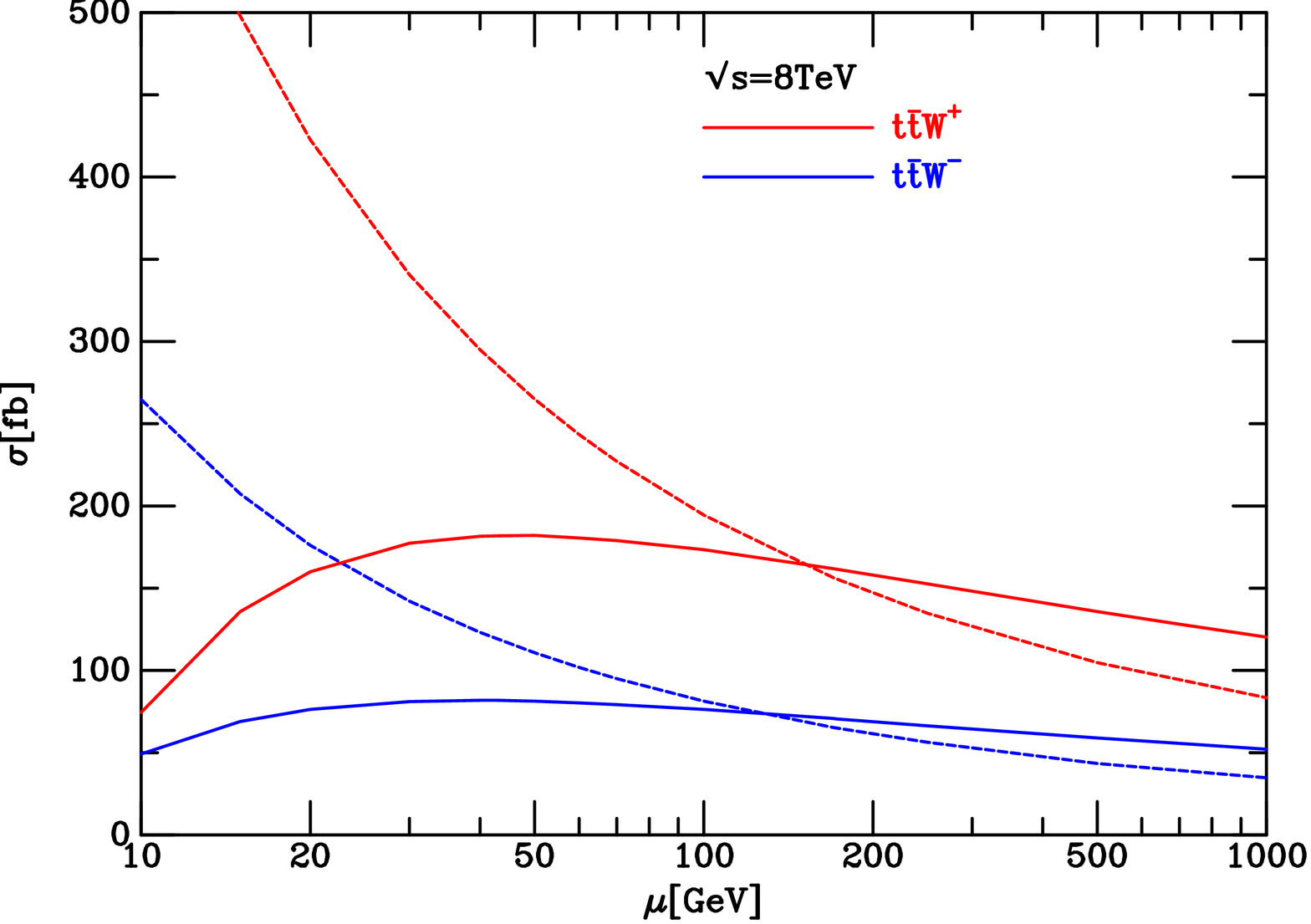}     
\caption{As in Fig.~\ref{ttW7} but at $\sqrt{s}=8$~TeV.}
\label{ttW8}
\end{center}
\end{figure}

\begin{figure}
\begin{center}
\includegraphics[angle=270,width=10cm]{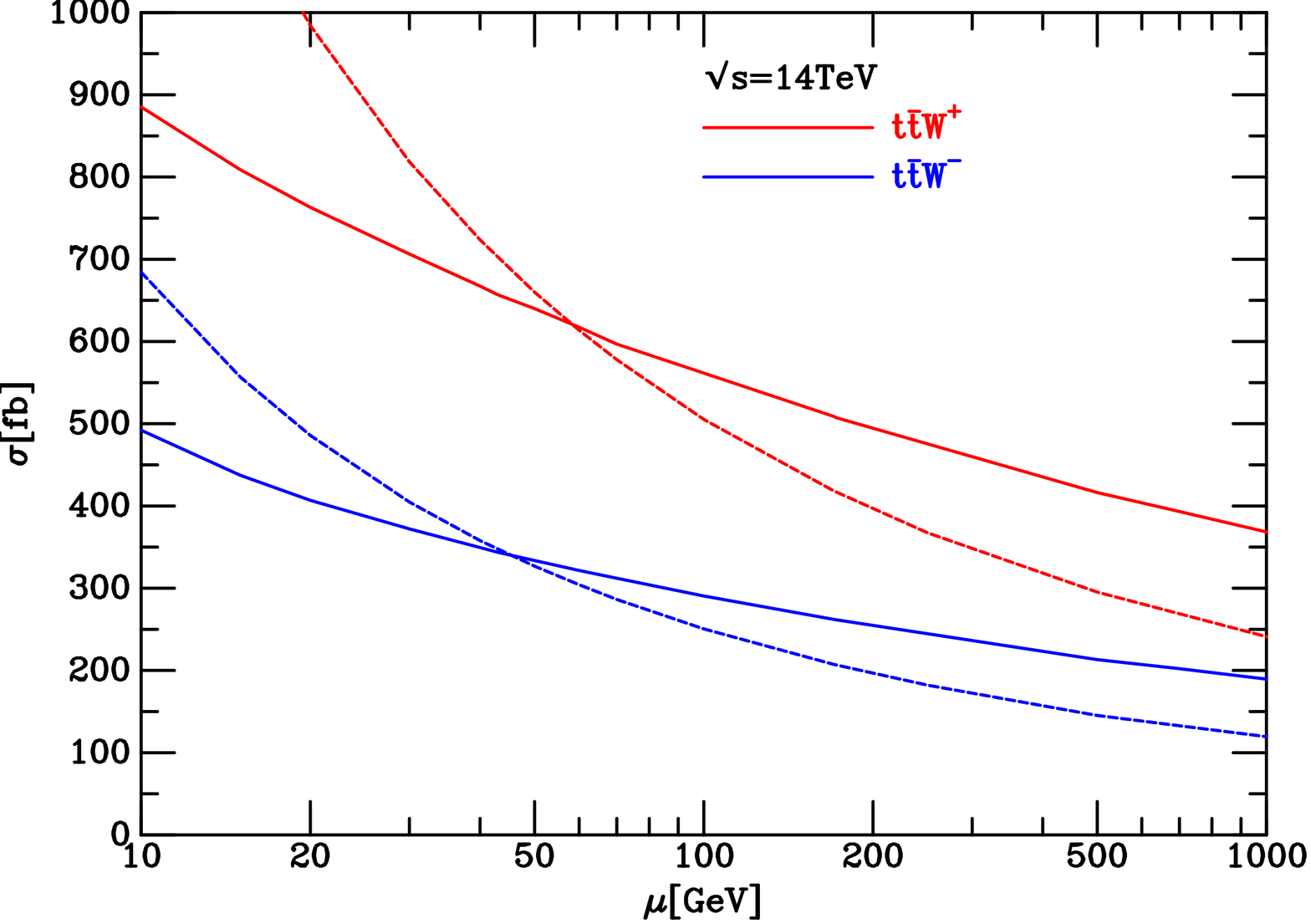}     
\caption{As in Fig.~\ref{ttW7} but at $\sqrt{s}=14$~TeV.}
\label{ttW14}
\end{center}
\end{figure}

We begin by assessing the scale dependence of the predictions at LO and NLO. This is particularly interesting
since one might imagine a rather wide possible choice of scales, say from $m_t$ to $2 m_t + m_W$.
Our results for the LHC operating at $\sqrt{s}=7$, $8$ and $14$~TeV, 
(setting the renormalization scale $\mu_r$ and factorization scale $\mu_f$ 
equal to a common scale $\mu$),
are shown in Figs.~\ref{ttW7}, \ref{ttW8} and~\ref{ttW14} respectively. 
As expected, the scale dependence is reduced after including the NLO effects and the behavior of the
predictions for $t\tb\, W^+$ and $t\tb\, W^-$ is very similar. Qualitatively the scale dependence is also very similar
at $\sqrt{s}=7$ and $8$~TeV.  However, at $14$~TeV the
contribution to the cross section from quark-gluon initial states, that enter the calculation for the
first time at $O(\alpha_s^3)$, is much more important. 
Since this is a LO contribution the inherent scale dependence
is uncancelled. As a result the improvement in scale dependence from LO to
NLO is less dramatic at $14$~TeV.

We now present our best predictions for the cross sections for this process at LO and NLO, together with an assessment of the
theoretical uncertainty at each order.  Our predictions at the various LHC operating energies are given in
Table~\ref{table:NLOrates}. The central value of the cross sections corresponds to a scale $\mu_r = \mu_f = m_t$ and the central
pdf set. The first uncertainty corresponds to a variation of the scales in the range $[m_t/4, 4m_t]$, which spans most of the
conceivably relevant kinematic region. The second uncertainty originates from the pdf fitting procedure and is computed with the
90\% confidence-level pdf sets and procedure of Ref.~\cite{Martin:2009iq}.  At NLO, where uncertainty sets are available
that include both pdf and $\alpha_s$ uncertainties, we use the method of Ref.~\cite{Martin:2009bu} to also include the 90\% confidence
level uncertainty in $\alpha_s(M_Z)$ (corresponding to $0.1163 < \alpha_s(M_Z) < 0.1233$).
Even considering the generous 90\% confidence-level ranges we have considered, the uncertainty from the pdf and $\alpha_s$
determination is smaller than the residual scale uncertainty of the NLO
calculation. At $14$~TeV this combined uncertainty is a factor of three smaller
than the uncertainty from the scale dependence. 
In summary, although the cross section has a smaller uncertainty at NLO than
at LO, the combined theoretical uncertainty is still sizeable (at best in the range of $30$-$40$\%) and grows with energy.
\renewcommand{\baselinestretch}{1.6}
\begin{table}
  \centering
\begin{tabular}{|c|c|c|c|}
\hline

\hline
{$t \tb\,  W^+$}  & $\sqrt{s}=7$~ TeV & $\sqrt{s}=8$~ TeV & $\sqrt{s}=14$~ TeV  \\
\hline
LO          &  $118^{+87\%}_{-40\%}$(scale)${}^{+6\%}_{-6\%}$(pdf)~~~~~
            &  $156^{+83\%}_{-39\%}$(scale)${}^{+6\%}_{-5\%}$(pdf)~~~~~
	    &  $416^{+68\%}_{-36\%}$(scale)${}^{+4\%}_{-4\%}$(pdf)~~~~~  \\
NLO         &  $119^{+8\%}_{-20\%} $(scale)${}^{+7\%}_{-8\%}$(pdf+$\alpha_s$)
            &  $161^{+12\%}_{-20\%}$(scale)${}^{+7\%}_{-8\%}$(pdf+$\alpha_s$)
	    &  $507^{+29\%}_{-22\%}$(scale)${}^{+7\%}_{-8\%}$(pdf+$\alpha_s$)  \\
\hline
\hline
{$t \tb\,  W^-$} & $\sqrt{s}=7$~ TeV & $\sqrt{s}=8$~ TeV & $\sqrt{s}=14$~ TeV  \\
\hline
LO          & $47^{+87\%}_{-41\%} $(scale)${}^{+6\%}_{-6\%}$(pdf)~~~~~
            & $65^{+84\%}_{-40\%} $(scale)${}^{+5\%}_{-6\%}$(pdf)~~~~~
            & $206^{+68\%}_{-36\%}$(scale)${}^{+4\%}_{-5\%}$(pdf)~~~~~  \\
NLO         & $50^{+12\%}_{-21\%} $(scale)${}^{+6\%}_{-8\%}$(pdf+$\alpha_s$)
            & $71^{+16\%}_{-21\%} $(scale)${}^{+6\%}_{-8\%}$(pdf+$\alpha_s$)
	    & $262^{+31\%}_{-23\%}$(scale)${}^{+7\%}_{-8\%}$(pdf+$\alpha_s$) \\
\hline
\end{tabular}
\renewcommand{\baselinestretch}{1}
\caption{Leading and NLO cross sections in femtobarns, with relative theoretical uncertainties
estimated from scale dependence (in the range $\mu =m_t/4, m_t, 4 m_t$) and 90\% confidence
pdf uncertainty sets (including also $\alpha_s$ variation at NLO).}
  \label{table:NLOrates}
\end{table}
\renewcommand{\baselinestretch}{1}

\section{$t \tb  \, W$ production including decay}
\begin{figure}
\begin{center}
\includegraphics[angle=270,width=7cm]{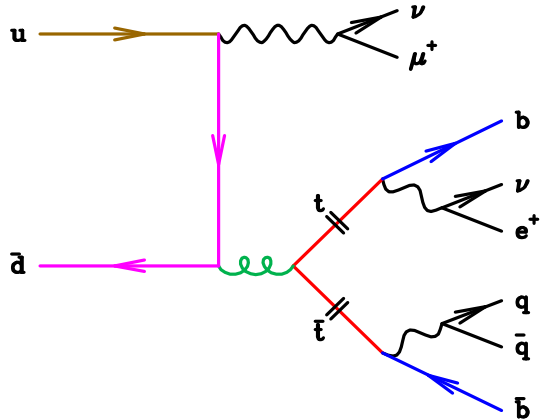}     
\caption{Example of a leading order diagram leading to a same-sign lepton event.
The double lines indicate that the top and anti-top are on mass shell.}
\label{ttbarw}
\end{center}
\end{figure}

One of the advantages of the amplitude method is that it allows inclusion of the decay of the top quark and subsequent
decay of the $W$-boson, with all the correct spin correlations included. An example of a lowest order diagram, that displays the full decay
chain that we include, is shown in Fig.~\ref{ttbarw}.
The inclusion of these decays is achieved with very
little computational cost due to the fact that we can factorize all amplitudes into production and decay
stages, see for example ref.~\cite{Campbell:2012uf}.
In view of the smallness of the cross section we have not yet implemented radiative corrections in the decay of the
top quark, although such an extension would be straightforward.


Our calculation can also be used to assess the effects of radiative corrections in kinematic regions
that are relevant for new physics searches. We shall study this in the following section, focussing in particular
on same-sign lepton searches.

\section{Same-sign lepton events}
As can be seen from Fig.~\ref{ttbarw}, the $t\tb\, W$ process can lead to final states that contain
same-sign leptons, a relatively rare phenomenon in the Standard Model. As such, these final states
form the basis of many searches for theories of new physics, in particular supersymmetry, that could give rise
to much higher rates. These searches can focus solely on the presence of two same-sign dileptons~\cite{Aad:2012cg} or,
more often, also require missing transverse momentum and jets~\cite{Chatrchyan:2011wba,Aad:2012xxx,CMS-PAS-SUS-11-010}.
The jets that are present may also be $b$-tagged~\cite{CMS-PAS-SUS-11-020}.

Same-sign lepton events originating from the Standard Model are experimentally of three
types. The first category is fakes,
for instance when a hadron is misidentified as a lepton. The second, ``q-flips'' occurs when
a lepton charge is mis-assigned so that an opposite-sign event enters the same-sign sample.
Both of these categories are difficult to assess theoretically and must be estimated by
data-driven techniques that rely on detailed detector simulation. The final category is
comprised of rare Standard Model background processes that, in general, cannot be measured with current
data samples. As a result, at present same-sign dilepton analyses must rely on theoretical
calculations for an assessment of this form of background.

Examples of rare Standard Model backgrounds are:
\begin{enumerate}
\item $W^+Z$, $W^-Z$ and $ZZ$ production, with $W^\pm \to \ell^\pm \nu$, $Z \to \ell^- \ell^+$. These processes can be computed
at NLO accuracy including all spin correlations, see for instance Refs.~\cite{Campbell:1999ah,Campbell:2011bn}. Predictions
at NLO, interfaced with a parton shower framework, are also available~\cite{Frixione:2007zp,Melia:2011tj}.
\item $W^\pm W^\pm$+dijets, with $W^\pm \to \ell^\pm \nu$. This process has been computed at NLO and included in the
POWHEG framework~\cite{Melia:2011gk}.
\item $W^\pm W^\pm$ production in vector boson fusion, with $W^\pm \to \ell^\pm \nu$. The NLO corrections
to this process were presented in Ref.~\cite{Jager:2009xx}.
\item $WWW$, $W^+W^-Z$, $ZZZ$ in various decay channels, producing same-sign leptons with either additional leptons or jets.
Predictions for these processes are available at NLO~\cite{Campanario:2008yg,Lazopoulos:2007ix}.
\item $t \tb\, Z$ production, with at least one of the top quarks decaying semi-leptonically. 
This cross-section is available at the NLO level~\cite{Lazopoulos:2007bv,Lazopoulos:2008de,Kardos:2011na,Garzelli:2011is},
but unlike our $t \tb\, W$ calculation the complete spin correlations are not included at NLO.
\item $t \tb\, W$ production, with one top quark decaying semi-leptonically and the other hadronically. 
The prediction of $t \tb\, W$ processes at NLO with subsequent decays (such as the one in the previous sentence) 
is the subject of this paper.
\end{enumerate}
The relative size of these various backgrounds is dictated by the further event selections that are applied
in the search. If the signal region is defined by the presence of additional jets, for instance, then the diboson
backgrounds shown above are relatively suppressed. Requiring that such jets are $b$-tagged favors both of the associated
top production processes, where two $b$-quarks are naturally produced.  


\begin{figure}[ht]
\begin{center}
\includegraphics[angle=0,width=7.5cm]{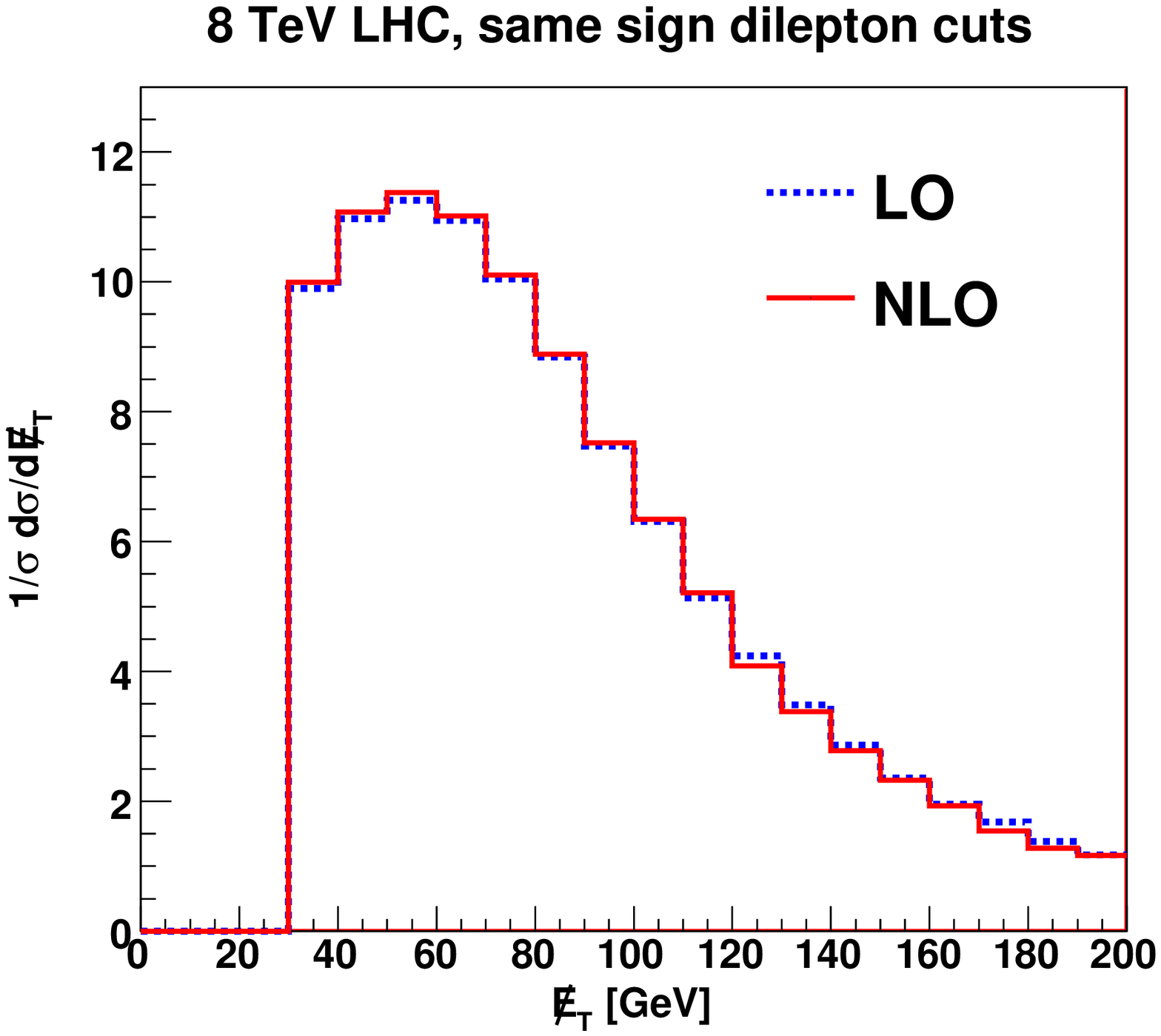}
\includegraphics[angle=0,width=7.5cm]{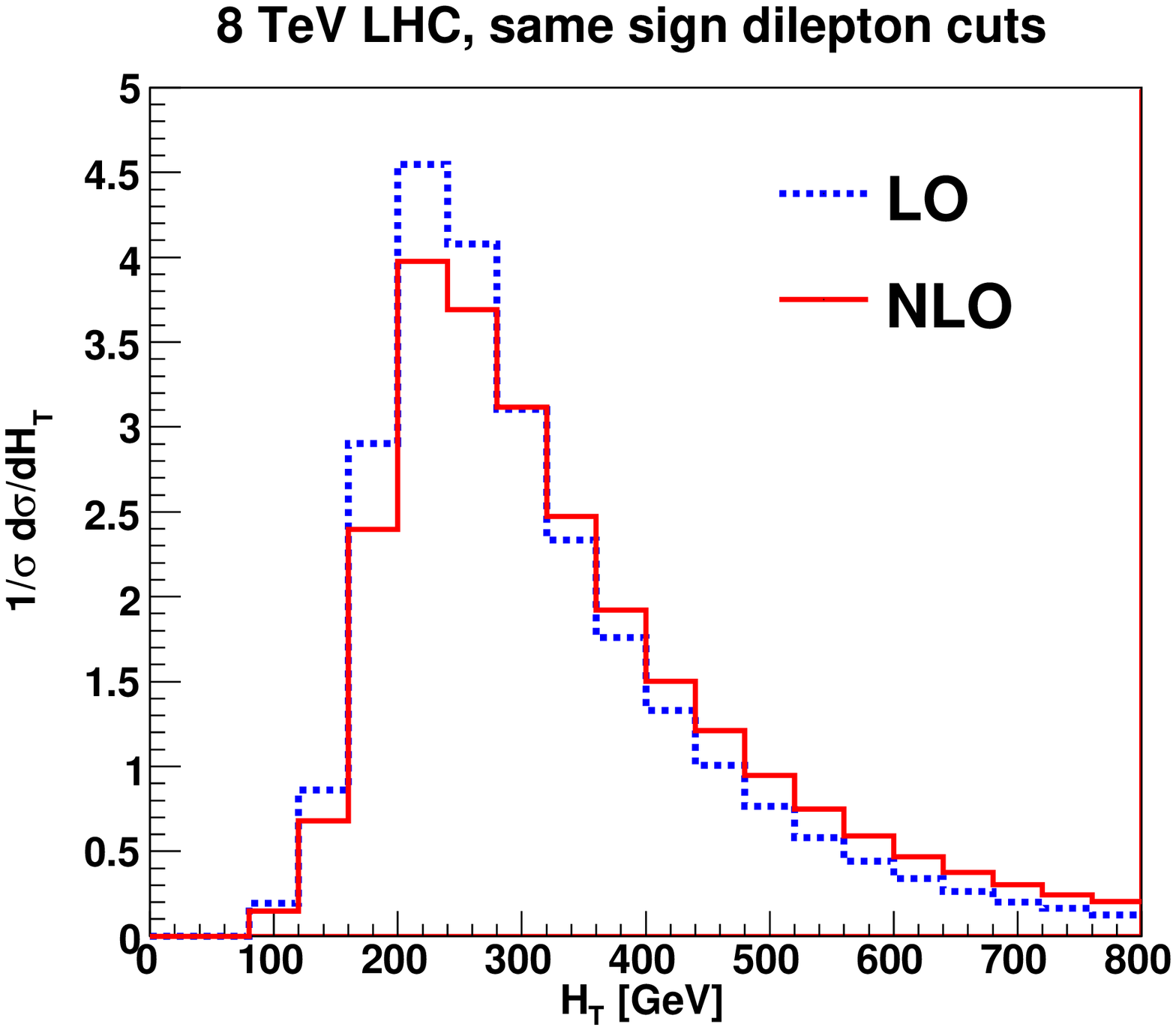}
\caption{The distributions of $\slsh{E}_T$ (left) and $H_T$ (right) for the $t\tb W^+$ process, with 
the top quark decaying leptonically and the antitop quark hadronically. The calculation is performed
for the $8$~TeV LHC, under the same-sign dilepton cuts described in the text. The LO prediction is the dashed (blue) histogram
while the NLO prediction is solid (red).}
\label{fig:METandHT}
\end{center}
\end{figure}
As an example that is especially sensitive to the $t \tb\, W$ Standard Model process, we focus on a recent CMS analysis that
requires the presence of same-sign dileptons, missing transverse momentum and $b$-tagged jets~\cite{CMS-PAS-SUS-11-020}.
We will provide predictions  for the $t\tb\, W$ background, with one top quark decaying leptonically and the other hadronically,
that would be relevant for this search if it were repeated
at $8$~TeV. Adopting the cuts from Ref.~\cite{CMS-PAS-SUS-11-020}, we require two same-sign leptons in the acceptance range,
\beq
p_T(\mbox{lepton}) > 20~\mbox{GeV} \;, \qquad |\eta(\mbox{lepton})| < 2.4 \;.
\eeq
Jets are clustered according to the anti-$k_T$ algorithm with a distance parameter
$D=0.5$ and we require that the algorithm finds at least two jets that satisfy,
\beq
p_T(\mbox{jet}) > 40~\mbox{GeV} \;, \qquad |\eta(\mbox{jet})| < 2.5 \;.
\eeq
Two of these jets should contain the $b$ and $\bar b$ quarks.
Finally, we require that the missing transverse momentum, which
in our calculation is given by the magnitude of the vector sum of the two neutrino
four-momenta, is greater than $30$~GeV.

We follow the division of events into signal regions that is presented in Ref.~\cite{CMS-PAS-SUS-11-020}.
These regions are defined by minimum cuts on the missing transverse momentum and the variable $H_T$ that is
defined by,
\beq
H_T = \sum_i p_T({\rm jet}~i) \;,
\eeq
i.e. it is the scalar sum of the jet transverse momenta.
Since these observables are crucial in the definition of the signal regions, in Fig.~\ref{fig:METandHT} we
present our predictions for them obtained at LO and NLO. These predictions are for a pair of positively-charged
leptons, i.e. the $t\tb\, W^+$ process, although they do not differ greatly for the negatively-charged pair ($t \tb\, W^-$).
We see that, although the shape of the missing transverse momentum distribution
is unaffected by radiative corrections, the shape of the $H_T$ distribution
is changed considerably. At NLO the peak around $250$~GeV is reduced but the
tail is somewhat higher.

\begin{figure}[ht]
\begin{center}
\includegraphics[angle=0,width=11.5cm]{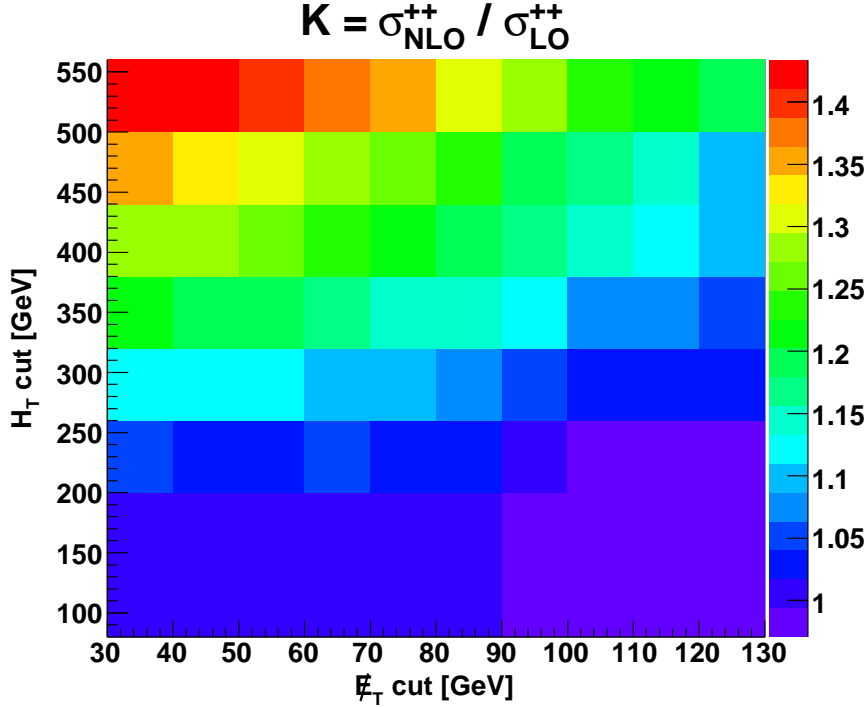}
\caption{The dependence of the $K$-factor for the positively-charged dilepton process on the cuts that are applied
to $\slsh{E}_T$ and $H_T$. The $K$-factor is defined for $\mu=m_t$, }
\label{fig:Kfac}
\end{center}
\end{figure}

\renewcommand{\baselinestretch}{1.6}
\begin{table}
\begin{center}
\begin{tabular}{|l|c|c|c|c|c|}
\hline
Signal region          & SR1/SR2 & SR3 & SR4 & SR5 & SR6 \\
\hline
$\slsh{E}_T$ cut [GeV] & $30$ & $120$ & $50$  & $50$  & $120$ \\
$H_T$ cut [GeV]        & $80$ & $200$ & $200$ & $320$ & $320$ \\
\hline
$\sigma_{LO}^{++}$  & $0.42^{+83\%}_{-40\%}{}^{+ 6\%}_{- 5\%}$
                    & $0.098^{+93\%}_{-42\%}{}^{+ 7\%}_{- 6\%}$
		    & $0.29^{+83\%}_{-41\%}{}^{+ 6\%}_{- 6\%}$ 
                    & $0.14^{+86\%}_{-43\%}{}^{+ 7\%}_{- 6\%}$ 
		    & $0.064^{+98\%}_{-42\%}{}^{+ 8\%}_{- 7\%} $ \\

$\sigma_{NLO}^{++}$ & $0.42^{+7\%}_{-19\%}{}^{+ 7\%}_{- 8\%}$
                    & $0.096^{+6\%}_{-22\%}{}^{+ 4\%}_{- 9\%}$
		    & $0.30^{+10\%}_{-20\%}{}^{+ 7\%}_{- 8\%}$ 
                    & $0.16^{+31\%}_{-25\%}{}^{+ 7\%}_{- 8\%}$
		    & $0.068^{+18\%}_{-26\%}{}^{+ 5\%}_{-10\%}$ \\
\hline
$\sigma_{LO}^{--}$  & $0.17^{+88\%}_{-35\%}{}^{+ 6\%}_{- 6\%}$
                    & $0.037^{+92\%}_{-41\%}{}^{+ 7\%}_{- 7\%}$
		    & $0.12^{+83\%}_{-42\%}{}^{+ 6\%}_{- 6\%}$
                    & $0.051^{+94\%}_{-41\%}{}^{+ 7\%}_{- 7\%}$
		    & $0.023^{+96\%}_{-43\%}{}^{+ 8\%}_{- 8\%}$ \\

$\sigma_{NLO}^{--}$ & $0.19^{+11\%}_{-21\%}{}^{+ 6\%}_{- 8\%}$
                    & $0.038^{+8\%}_{-21\%}{}^{+ 7\%}_{- 8\%}$
		    & $0.13^{+15\%}_{-23\%}{}^{+ 7\%}_{- 8\%}$              
                    & $0.067^{+42\%}_{-28\%}{}^{+ 7\%}_{- 8\%}$
		    & $0.025^{+24\%}_{-24\%}{}^{+ 7\%}_{- 7\%}$ \\
\hline
\end{tabular}
\renewcommand{\baselinestretch}{1}
\caption{Definition of signal regions and cross sections in femtobarns for a single
flavor of lepton in each leptonic $W$ decay. The first quoted uncertainty corresponds to variation of the scale by a factor
of four in each direction about the central value, $\mu = m_t$. The second range
corresponds to the $90$\% confidence level pdf uncertainty (LO) or pdf+$\alpha_s$ uncertainty (NLO).} 
\label{table:SRtable}
\end{center}
\end{table}
\renewcommand{\baselinestretch}{1}
Our results for the various signal regions at $8$~TeV are presented in Table~\ref{table:SRtable}. Note that we show
predictions for positively- and negatively-charged lepton pairs separately and not the sum. As such, regions
SR1 and SR2 of Ref.~\cite{CMS-PAS-SUS-11-020} are conflated. Moreover, our process contains only two $b$-quarks
and so does not contribution to the region SR7.
The change in cross section from LO to NLO in each of the signal regions is relatively mild
but the reduction in theoretical uncertainty from the choice of scale is substantial.
Both the change in cross section and the associated scale uncertainty are largest for SR5, one of the
regions defined by the largest $H_T$ cut. Since the $\slsh{E}_T$ cut is relatively small for this
region, these results can be explained by the fact that the calculation is
more susceptible to the change in shape of the $H_T$ distribution shown in Fig.~\ref{fig:METandHT}.
We explore the effect of the NLO corrections to the positively-charged dilepton process more generally in Fig.~\ref{fig:Kfac}.
The figure shows the $K$-factor, i.e. the ratio of the NLO and LO predictions,
as a function of the $\slsh{E}_T$ and $H_T$ cuts in the range $30$--$130$~GeV ($\slsh{E}_T$) and $80$--$500$~GeV ($H_T$).
The $K$-factor is evaluated at our central scale choice, $\mu_r=\mu_f=m_t$.
For most of these choices the $K$-factor is close to unity, with the value rising as the $H_T$ cut
is raised and the $\slsh{E}_T$ cut is reduced.

\section{Conclusions}
We have presented first NLO results on the process $t\tb\, W^\pm$ including the decay of the
top quark and the vector bosons. The cross section has interest, both as a Standard Model
measurement, and as a source of events that contain same-sign leptons,
missing energy, jets and $b$-quarks. 

At all potential LHC operating energies, between $7$ and
$8$~TeV, evaluating cross sections with a central scale choice of $\mu = m_t$ leads to very
little enhancement of the LO cross section at NLO. For a large scale choice, such as $\mu=2m_t+m_W$
the $K$-factor is in the region $1.3$--$1.4$. Considering scale variation and the combined
pdf and $\alpha_s$ uncertainty, the overall accuracy of the NLO prediction is at the level of
$30$\% at best.

At NLO the $H_T$ observable, defined as the scalar sum of all jet transverse momenta in the
event, is subject to important NLO corrections. These cause more events to be found at high
$H_T$ so that NLO effects are more pronounced as progressively harder cuts on $H_T$ are performed.
This is of particular interest in the same-sign dilepton channel, where cuts of exactly this
nature are usually performed in order to maximize the sensitivity to new physics.

\medskip

\noindent
{\bf \large Acknowledgments} 

We gratefully acknowledge useful conversations with Frank Petriello.
This research is supported by the US DOE under contract
DE-AC02-06CH11357.
    
\bibliography{ttw}

\providecommand{\href}[2]{#2}\begingroup\raggedright\begin{thebibliography}{10}

\bibitem{Badger:2010mg}
S.~Badger, J.~M. Campbell, and R.~Ellis, {\it {QCD corrections to the hadronic
  production of a heavy quark pair and a W-boson including decay
  correlations}},  {\em JHEP} {\bf 1103} (2011) 027,
  [\href{http://xxx.lanl.gov/abs/1011.6647}{{\tt arXiv:1011.6647}}].

\bibitem{Campbell:2004ch}
J.~M. Campbell, R.~K. Ellis, and F.~Tramontano, {\it {Single top production and
  decay at next-to-leading order}},  {\em Phys.Rev.} {\bf D70} (2004) 094012,
  [\href{http://xxx.lanl.gov/abs/hep-ph/0408158}{{\tt hep-ph/0408158}}].

\bibitem{Campbell:2012uf}
J.~M. Campbell and R.~K. Ellis, {\it {Top-quark processes at NLO in production
  and decay}},  \href{http://xxx.lanl.gov/abs/1204.1513}{{\tt
  arXiv:1204.1513}}.

\bibitem{Beenakker:2002nc}
W.~Beenakker, S.~Dittmaier, M.~Kramer, B.~Plumper, M.~Spira, et~al., {\it {NLO
  QCD corrections to t anti-t H production in hadron collisions}},  {\em
  Nucl.Phys.} {\bf B653} (2003) 151--203,
  [\href{http://xxx.lanl.gov/abs/hep-ph/0211352}{{\tt hep-ph/0211352}}].

\bibitem{Dawson:2002tg}
S.~Dawson, L.~Orr, L.~Reina, and D.~Wackeroth, {\it {Associated top quark Higgs
  boson production at the LHC}},  {\em Phys.Rev.} {\bf D67} (2003) 071503,
  [\href{http://xxx.lanl.gov/abs/hep-ph/0211438}{{\tt hep-ph/0211438}}].

\bibitem{Dawson:2003zu}
S.~Dawson, C.~Jackson, L.~Orr, L.~Reina, and D.~Wackeroth, {\it {Associated
  Higgs production with top quarks at the large hadron collider: NLO QCD
  corrections}},  {\em Phys.Rev.} {\bf D68} (2003) 034022,
  [\href{http://xxx.lanl.gov/abs/hep-ph/0305087}{{\tt hep-ph/0305087}}].

\bibitem{Lazopoulos:2007bv}
A.~Lazopoulos, K.~Melnikov, and F.~J. Petriello, {\it {NLO QCD corrections to
  the production of $t \bar{t} Z$ in gluon fusion}},  {\em Phys.Rev.} {\bf D77}
  (2008) 034021, [\href{http://xxx.lanl.gov/abs/0709.4044}{{\tt
  arXiv:0709.4044}}].

\bibitem{Lazopoulos:2008de}
A.~Lazopoulos, T.~McElmurry, K.~Melnikov, and F.~Petriello, {\it
  {Next-to-leading order QCD corrections to $t \bar{t} Z$ production at the
  LHC}},  {\em Phys.Lett.} {\bf B666} (2008) 62--65,
  [\href{http://xxx.lanl.gov/abs/0804.2220}{{\tt arXiv:0804.2220}}].

\bibitem{Kardos:2011na}
A.~Kardos, Z.~Trocsanyi, and C.~Papadopoulos, {\it {Top quark pair production
  in association with a Z-boson at NLO accuracy}},  {\em Phys.Rev.} {\bf D85}
  (2012) 054015, [\href{http://xxx.lanl.gov/abs/1111.0610}{{\tt
  arXiv:1111.0610}}].

\bibitem{Garzelli:2011is}
M.~Garzelli, A.~Kardos, C.~Papadopoulos, and Z.~Trocsanyi, {\it {Z0 - boson
  production in association with a top anti-top pair at NLO accuracy with
  parton shower effects}},  \href{http://xxx.lanl.gov/abs/1111.1444}{{\tt
  arXiv:1111.1444}}.

\bibitem{Melnikov:2011qx}
K.~Melnikov, A.~Scharf, and M.~Schulze, {\it {Top quark pair production in
  association with a jet: QCD corrections and jet radiation in top quark
  decays}},  {\em Phys.Rev.} {\bf D85} (2012) 054002,
  [\href{http://xxx.lanl.gov/abs/1111.4991}{{\tt arXiv:1111.4991}}].

\bibitem{Melnikov:2011ta}
K.~Melnikov, M.~Schulze, and A.~Scharf, {\it {QCD corrections to top quark pair
  production in association with a photon at hadron colliders}},  {\em
  Phys.Rev.} {\bf D83} (2011) 074013,
  [\href{http://xxx.lanl.gov/abs/1102.1967}{{\tt arXiv:1102.1967}}].

\bibitem{Bredenstein:2008zb}
A.~Bredenstein, A.~Denner, S.~Dittmaier, and S.~Pozzorini, {\it {NLO QCD
  corrections to t anti-t b anti-b production at the LHC: 1. Quark-antiquark
  annihilation}},  {\em JHEP} {\bf 0808} (2008) 108,
  [\href{http://xxx.lanl.gov/abs/0807.1248}{{\tt arXiv:0807.1248}}].

\bibitem{Bredenstein:2009aj}
A.~Bredenstein, A.~Denner, S.~Dittmaier, and S.~Pozzorini, {\it {NLO QCD
  corrections to pp $\to$ t anti-t b anti-b + X at the LHC}},  {\em
  Phys.Rev.Lett.} {\bf 103} (2009) 012002,
  [\href{http://xxx.lanl.gov/abs/0905.0110}{{\tt arXiv:0905.0110}}].

\bibitem{Bevilacqua:2009zn}
G.~Bevilacqua, M.~Czakon, C.~Papadopoulos, R.~Pittau, and M.~Worek, {\it
  {Assault on the NLO Wishlist: pp $\to$ t anti-t b anti-b}},  {\em JHEP} {\bf
  0909} (2009) 109, [\href{http://xxx.lanl.gov/abs/0907.4723}{{\tt
  arXiv:0907.4723}}].

\bibitem{Bredenstein:2010rs}
A.~Bredenstein, A.~Denner, S.~Dittmaier, and S.~Pozzorini, {\it {NLO QCD
  Corrections to Top Anti-Top Bottom Anti-Bottom Production at the LHC: 2. full
  hadronic results}},  {\em JHEP} {\bf 1003} (2010) 021,
  [\href{http://xxx.lanl.gov/abs/1001.4006}{{\tt arXiv:1001.4006}}].

\bibitem{Hirschi:2011pa}
V.~Hirschi, R.~Frederix, S.~Frixione, M.~V. Garzelli, F.~Maltoni, et~al., {\it
  {Automation of one-loop QCD corrections}},  {\em JHEP} {\bf 1105} (2011) 044,
  [\href{http://xxx.lanl.gov/abs/1103.0621}{{\tt arXiv:1103.0621}}].

\bibitem{Martin:2009iq}
A.~D. Martin, W.~J. Stirling, R.~S. Thorne, and G.~Watt, {\it {Parton
  distributions for the LHC}},  {\em Eur. Phys. J.} {\bf C63} (2009) 189--285,
  [\href{http://xxx.lanl.gov/abs/0901.0002}{{\tt arXiv:0901.0002}}].

\bibitem{Martin:2009bu}
A.~Martin, W.~Stirling, R.~Thorne, and G.~Watt, {\it {Uncertainties on alpha(S)
  in global PDF analyses and implications for predicted hadronic cross
  sections}},  {\em Eur.Phys.J.} {\bf C64} (2009) 653--680,
  [\href{http://xxx.lanl.gov/abs/0905.3531}{{\tt arXiv:0905.3531}}].

\bibitem{Aad:2012cg}
{\bf ATLAS} Collaboration, G.~Aad et~al., {\it {Search for anomalous production
  of prompt like-sign muon pairs and constraints on physics beyond the Standard
  Model with the ATLAS detector}},  {\em Phys.Rev.} {\bf D88} (2012) 032004,
  [\href{http://xxx.lanl.gov/abs/1201.1091}{{\tt arXiv:1201.1091}}].

\bibitem{Chatrchyan:2011wba}
{\bf CMS} Collaboration, S.~Chatrchyan et~al., {\it {Search for new physics
  with same-sign isolated dilepton events with jets and missing transverse
  energy at the LHC}},  {\em JHEP} {\bf 1106} (2011) 077,
  [\href{http://xxx.lanl.gov/abs/1104.3168}{{\tt arXiv:1104.3168}}].

\bibitem{Aad:2012xxx}
{\bf ATLAS} Collaboration, G.~Aad et~al., {\it {Search for gluinos in events
  with two same-sign leptons, jets and missing transverse momentum with the
  ATLAS detector in pp collisions at sqrt(s) = 7 TeV}},
  \href{http://xxx.lanl.gov/abs/1203.5763}{{\tt arXiv:1203.5763}}.

\bibitem{CMS-PAS-SUS-11-010}
{\bf CMS} Collaboration, {\it Search for new physics with same-sign isolated
  dilepton events with jets and missing energy},  2012.
\newblock CMS-PAS-SUS-11-010.

\bibitem{CMS-PAS-SUS-11-020}
{\bf CMS} Collaboration, {\it Search for new physics in events with same-sign
  dileptons, b-tagged jets and missing energy},  2012.
\newblock \url{http://cdsweb.cern.ch/record/1434376}.

\bibitem{Campbell:1999ah}
J.~M. Campbell and R.~K. Ellis, {\it {An update on vector boson pair production
  at hadron colliders}},  {\em Phys. Rev.} {\bf D60} (1999) 113006,
  [\href{http://xxx.lanl.gov/abs/hep-ph/9905386}{{\tt hep-ph/9905386}}].

\bibitem{Campbell:2011bn}
J.~M. Campbell, R.~K. Ellis, and C.~Williams, {\it {Vector boson pair
  production at the LHC}},  {\em JHEP} {\bf 1107} (2011) 018,
  [\href{http://xxx.lanl.gov/abs/1105.0020}{{\tt arXiv:1105.0020}}].

\bibitem{Frixione:2007zp}
S.~Frixione, E.~Laenen, P.~Motylinski, and B.~R. Webber, {\it {Angular
  correlations of lepton pairs from vector boson and top quark decays in Monte
  Carlo simulations}},  {\em JHEP} {\bf 0704} (2007) 081,
  [\href{http://xxx.lanl.gov/abs/hep-ph/0702198}{{\tt hep-ph/0702198}}].

\bibitem{Melia:2011tj}
T.~Melia, P.~Nason, R.~Rontsch, and G.~Zanderighi, {\it {$W^+W^-, WZ$ and $ZZ$
  production in the POWHEG BOX}},  {\em JHEP} {\bf 1111} (2011) 078,
  [\href{http://xxx.lanl.gov/abs/1107.5051}{{\tt arXiv:1107.5051}}].

\bibitem{Melia:2011gk}
T.~Melia, P.~Nason, R.~Rontsch, and G.~Zanderighi, {\it {$W^+W^+$ plus dijet
  production in the POWHEGBOX}},  {\em Eur.Phys.J.} {\bf C71} (2011) 1670,
  [\href{http://xxx.lanl.gov/abs/1102.4846}{{\tt arXiv:1102.4846}}].

\bibitem{Jager:2009xx}
B.~Jager, C.~Oleari, and D.~Zeppenfeld, {\it {Next-to-leading order QCD
  corrections to $W^+ W^+$ jj and $W^- W^-$ jj production via weak-boson
  fusion}},  {\em Phys.Rev.} {\bf D80} (2009) 034022,
  [\href{http://xxx.lanl.gov/abs/0907.0580}{{\tt arXiv:0907.0580}}].

\bibitem{Campanario:2008yg}
F.~Campanario, V.~Hankele, C.~Oleari, S.~Prestel, and D.~Zeppenfeld, {\it {QCD
  corrections to charged triple vector boson production with leptonic decay}},
  {\em Phys.Rev.} {\bf D78} (2008) 094012,
  [\href{http://xxx.lanl.gov/abs/0809.0790}{{\tt arXiv:0809.0790}}].

\bibitem{Lazopoulos:2007ix}
A.~Lazopoulos, K.~Melnikov, and F.~Petriello, {\it {QCD corrections to
  tri-boson production}},  {\em Phys.Rev.} {\bf D76} (2007) 014001,
  [\href{http://xxx.lanl.gov/abs/hep-ph/0703273}{{\tt hep-ph/0703273}}].

\end{thebibliography}\endgroup
\bibliographystyle{JHEP}

\end{document}